\documentclass{JHEP3}

\usepackage{multicol}
\usepackage[dvips]{epsfig,graphics}
\usepackage{graphicx}

 \newlength{\wth}
 \newcommand{\twographs}[2]{%
 \unitlength=1.1in
 \begin{picture}(5.8,2.6)
 \put(0,0){\epsfig{file=#1.eps, width=\wth}}
 \put(2.7,0){\epsfig{file=#2.eps, width=\wth}}
 \put(0,2.1){(a)}
 \put(2.7,2.1){(b)}
 \end{picture}
}
 \newcommand{\fourgraphs}[4]{%
 \unitlength=1.1in
 \begin{picture}(6.,5)
 \put(0,0){\epsfig{file=#3, width=\wth}}
 \put(3.,0){\epsfig{file=#4, width=\wth}}
 \put(0,2.6){\epsfig{file=#1, width=\wth}}
 \put(3.,2.6){\epsfig{file=#2, width=\wth}}
 \put(0,2.1){(c)}
 \put(3.0,2.1){(d)}
 \put(0,4.7){(a)}
 \put(3.0,4.7){(b)}
 \end{picture}
}
\newcommand{\twographsg}[2]{%
 \unitlength=1in
 \begin{picture}(6.,2.6)
 \put(0,0){\epsfig{file=#1, width=0.9\wth}}
 \put(3.,0){\epsfig{file=#2, width=0.9\wth}}
 \put(0,2.1){(a)}
 \put(3.0,2.1){(b)}
 \end{picture}
}

\preprint{hep-ph/0502151\\DAMTP-2005-08\\DFPD-05/TH/10\\IFT-UAM/CSIC-05-12}
\author{B.C. Allanach\\
DAMTP, CMS, University of Cambridge, Wilberforce Rd, CB3 0WA, UK}
\author{A. Brignole\\
Dipartimento di Fisica ,`G. Galilei', Universit\`{a} di Padova and INFN,
Sezione
di Padova, Via Marzolo 8, I-35131 Padua, Italy}
\author{L.E. Ib\'a\~nez\\
Departamento de F\'{i}sica T\'{e}orica C-XI and Instituto de F\'{i}sica 
Te\'{o}rica
C-XVI, 
Universidad Aut\'{o}noma de Madrid, Cantoblanco, 28049 Madrid, Spain}
\abstract{We analyze the phenomenology of a set of 
minimal supersymmetric standard model (MSSM)
soft terms inspired by flux-induced supersymmetry (SUSY)-breaking
in Type IIB string orientifolds. 
The scheme is extremely constrained with essentially only two 
free mass parameters: a parameter $M$, which sets the scale
of soft terms, and the $\mu$ parameter.
After imposing consistent radiative electro-weak symmetry breaking (EWSB)
the model depends upon one mass parameter (say, $M$).
In spite of being so constrained one finds consistency
with EWSB conditions. 
We demonstrate that those conditions have two solutions
for $\mu<0$, and none for $\mu>0$. The parameter $\tan \beta$ results 
as a prediction and is approximately 3--5 for one solution, 
and 25--40 for the other, depending upon $M$ and the top mass. 
We examine further constraints on the  model  coming from 
$b \rightarrow s \gamma$, the muon $g-2$, Higgs mass
limits and WMAP constraints on  dark matter. 
The MSSM spectrum is predicted in terms of the single free 
parameter $M$. 
The low $\tan \beta$ branch is consistent with a 
relatively light spectrum although it is compatible with 
standard cosmology only if the lightest neutralino is unstable.
The high $\tan \beta$ branch is compatible with all phenomenological 
constraints, but has quite a heavy spectrum. We argue that
the fine-tuning associated to this heavy spectrum would be
substantially ameliorated if an additional  relationship $\mu=-2M$ 
were  present in the underlying theory.}

\title{Phenomenology of a Fluxed MSSM}

\begin{document}

\section{Introduction}

The MSSM is one of the most promising candidates for an extension of
the Standard Model (SM). In applications to phenomenology a crucial ingredient is that
of the structure of SUSY-breaking soft terms.
A large number of models and scenarios for  
soft terms have been proposed in the literature 
(see e.g. ref.\cite{gordy} for a recent review  and references). 
Some of the most popular schemes, like dilaton/modulus
dominance SUSY-breaking and their generalisations are
inspired by heterotic string models (see e.g.\cite{bim2} for
details and references). 

In this context it has been realised in the
last few years that strings other than the heterotic, 
particularly Type II and Type I string theories,  are 
equally viable as candidates in which to embed the MSSM.  
A first analysis of the soft terms
in this class of models was presented in ref.\cite{imr}.
More recently it has been found  that fluxes of
antisymmetric 3-forms which are present in Type IIB string
theory are  natural sources for SUSY-breaking
in Type IIB orientifold models
\cite{fluxsoft}. In the latter the SM
fields are assumed to live on the world-volume of
D7 or/and D3 branes and/or their intersections.
Specifically, if one assumes that the SM fields 
correspond to `geometric moduli' of D7-branes, 
it was pointed out in ref.\cite{iflux} that
a very simple set of SUSY-breaking soft terms 
appear if certain background fluxes are present\footnote{For  other structures of MSSM soft terms corresponding
to a different localisation of SM particles on the D-branes 
see \cite{lrs,kklw,fi}.}.
In fact these soft terms may be understood  as 
coming from a modulus-dominance scheme applied
to the particular case of Type IIB orientifolds, 
which leads to results different to those in the heterotic case.
Although no specific string model with these  characteristics has  been 
constructed, the structure of soft terms is so simple and
predictive that it is certainly worthwhile examining 
its phenomenological viability.

The  boundary conditions for the SUSY breaking terms are a subset of those of
the minimal supergravity (mSUGRA) model. All scalars have a common mass $m_0$, 
the gauginos have a common mass $M_{1/2}$ and the trilinear scalar couplings
are identical to each other (once divided by the corresponding Yukawa
coupling) and denoted $A_0$. 
In the mentioned scheme 
\cite{iflux}
such parameters are constrained as follows:
\begin{equation}
 M_{1/2} = M \; , \;\;\; A_0 = -3 M \; , \;\;\; m_0 = |M|
\label{sugraCond}
\end{equation}
where $M$ parametrises the overall SUSY breaking mass
scale in the model. In the simplest case $M$ coincides
with the gravitino mass. This would imply that the gravitino is not the
lightest supersymmetric particle, so it tends to decouple from
phenomenology, being very weakly coupled to matter. We should bear 
in mind, however, that the relationship between $M$ and the 
gravitino mass may vary in particular
models. Another independent mass parameter of the model is $\mu$,
which appears in the superpotential term $-\mu H_1 H_2$.
We parametrise the associated soft breaking term in the 
scalar potential as $-\mu B H_1 H_2$.  
Also the $B$ soft parameter is predicted in terms of $M$:
\begin{equation}
B = - 2 M
\label{bCond}
\end{equation}
One of the nice features of the set of soft terms in
eqs.~(\ref{sugraCond}), (\ref{bCond}) is that
complex phases may be rotated away, hence there is no
`SUSY-CP problem'. 
The MSSM with the above structure of
soft terms was named as the `fluxed MSSM' in \cite{iflux}
(where $\mu B$ was denoted as $-B$).

Note that the above  set of soft terms is 
extremely  restrictive. 
There are  only two free mass parameters
$\mu $ and $M$. Imposing
appropriate electroweak symmetry breaking (EWSB) essentially 
will lead us  to a single parameter (say, $M$) corresponding
to the overall scale of SUSY-breaking. Thus it
is not at all obvious that consistent radiative EWSB 
may be obtained in such a constrained system.
Remarkably, we find that the  above soft terms are
consistent with EWSB with $\tan \beta $ being fixed
around two regions, with $\tan \beta =3-5$  and
$\tan \beta = 25-40$ respectively. The complete sparticle 
spectrum is then fixed  depending only on $M$ for those
two regions. In what follows we will carry out a detailed analysis 
of the conditions of radiative EWSB in this scheme.
We will also present constraints coming from  $b\rightarrow s \gamma$,
the muon $g-2$, Higgs mass limits and dark matter relic 
density.

It is interesting to compare these results with those coming from
other string-inspired schemes with a reduced number of soft 
parameters. One of the most attractive schemes is that 
of the dilaton-domination scenario \cite{dilaton} in heterotic
string models in which the boundary conditions are given by  
$M_{1/2}= - A_0=\pm \sqrt{3}m_0$, $m_0$ being a universal scalar mass.
In this case there are  three free parameters corresponding to $m_0$,
$\mu $ and $B$, 
hence the general dilaton domination scenario is less predictive 
than the fluxed MSSM model here analysed which has only two free
parameters, $M$ and $\mu $, $B$ being fixed by eq.~(\ref{bCond}).
It is thus particularly remarkable that correct radiative 
EWSB may be achieved in this very constrained system. 
There are, however, some restricted versions of
the dilaton domination scenario which are equally constrained,
and proper EWSB can be obtained even there. 
This happens, for instance, if $\mu $ is generated through 
a Giudice-Masiero mechanism \cite{gm}, which leads to the
the further constraint $B=2m_0$ \cite{dilaton,blm}.
Proper EWSB can be obtained\protect\footnote{This holds if
$\mu$ is not specified. In an even more constrained scenario
considered in \cite{bimmu}, in which both $\mu$ and $B$ are 
predicted, dilaton dominance is not compatible with EWSB.}
provided $B$ and $M_{1/2}$ have opposite sign \cite{clm},
as in our case (see eq.~(\ref{bCond})). 

A few  comments are in order about our notation and
procedure. In general, we will  write the soft terms  
in the notation of {\small \tt SOFTSUSY} \cite{Allanach:2001kg} 
(except for $\mu B = m_3^2$).
We can take $M$ real and positive without loss of generality.
In order to discuss EWSB,
it is convenient to replace the boundary condition eq.~(\ref{bCond})
with a more general form, that is
\begin{equation}
B = - 2 M r_B .
\label{xtraCond}
\end{equation}
The default case corresponds to $r_B=1$. 
Eqs.~(\ref{sugraCond}), (\ref{xtraCond}) are to be applied at
some high fundamental scale. For simplicity, we take such a scale to be
$M_{GUT}$, which is defined by the scale of unification of the 
GUT-normalised gauge couplings
\begin{equation}
g_1(M_{GUT}) = g_2(M_{GUT}) \label{gaugeCond}.
\end{equation}
We have applied the constraints at the gauge unification scale as an 
approximation. In principle, one should apply the constraints at 
the fundamental string scale, which may or may not coincide with the gauge 
unification scale. In the case that the string scale is not too many 
orders of magnitude greater than the gauge unification scale, our 
approximation should hold quite well.
Later we will estimate this effect (or other possible
corrections)
by perturbing the fluxed boundary conditions imposed at $M_{GUT}$.
The strong gauge coupling is set from low energy data, and so $g_3(M_{GUT})
\neq g_{1,2} (M_{GUT})$ is assumed to be shifted to the unified value by 
GUT scale threshold effects. We have checked that the required correction
is at the per cent level only.

We will use the following default inputs for Standard Model
quantities, except 
where explicitly stated: the pole mass of the top quark~\cite{Abazov:2004cs}
$m_t=178$ GeV, the running Standard Model value of the bottom quark
mass~\cite{Eidelman:2004wy}
$m_b(m_b)^{\overline {MS}}_{SM}=4.25$ GeV, the strong coupling 
constant~\cite{Eidelman:2004wy} $\alpha_s(M_Z)=0.1187$, 
$\alpha^{-1}(M_Z)^{\overline {MS}}=127.918$~\cite{Eidelman:2004wy}
and
$M_Z=91.1187$ GeV~\cite{Eidelman:2004wy}. 

\section{Electroweak Symmetry Breaking}
The standard MSSM tree-level minimisation of the Higgs potential
leads to the equations
\begin{eqnarray}
\mu^2 & = & 
\frac{ - m_{H_2}^2\tan^2 \beta + m_{H_1}^2}{\tan^2 \beta-1} 
- \frac{1}{2} M_Z^2 ,
\label{mucond}
\\
\mu B & = &  \frac{1}{2} \sin 2 \beta \, (m_{H_1}^2 +m_{H_2}^2 + 2 \mu^2).
\label{Bcond}
\end{eqnarray}
As is well known, the minimisation conditions of the loop-corrected Higgs 
potential can also be cast in this form, with appropriate interpretations of 
the parameters (in particular, shifting $m_{H_i}^2$ through tadpole terms).
The minimisation conditions are imposed, as usual, at the
scale $M_{SUSY}=\sqrt{m_{{\tilde t}_1} m_{{\tilde t}_2}}$.
The default version of {\small \tt SOFTSUSY1.9} fixes 
$\tan \beta$ and $sgn(\mu)$ as input parameters and allows 
us to extract $|\mu(M_{SUSY})|$ and $B(M_{SUSY})$.
In this approach, which we will follow in this section only,
the value of $B(M_{GUT})$  (i.e. $r_B$ in eq.~(\ref{xtraCond}))
is computed rather than imposed. This will allow us to examine 
the consistency of electroweak symmetry breaking with 
the flux-induced soft terms.
More specifically, {\small \tt SOFTSUSY1.9} allows us to interpolate 
the parameters of the MSSM below the weak scale and the GUT scale 
by integrating 2-loop MSSM renormalisation group equations (RGEs). 
It solves these RGEs while simultaneously imposing the boundary
conditions eqs.~(\ref{sugraCond}), (\ref{gaugeCond}) and adjusting the
Yukawa and gauge couplings so that they agree with the data. For more
details, refer to the {\small \tt SOFTSUSY} manual~\cite{Allanach:2001kg}.

\FIGURE{\twographs{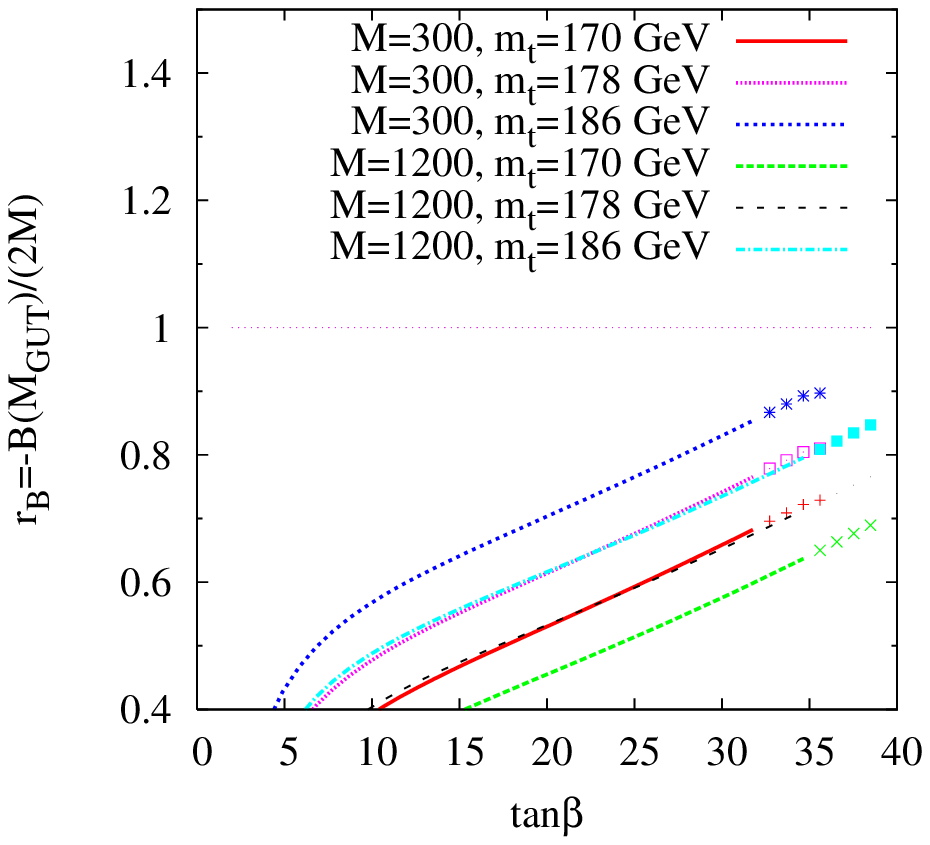}{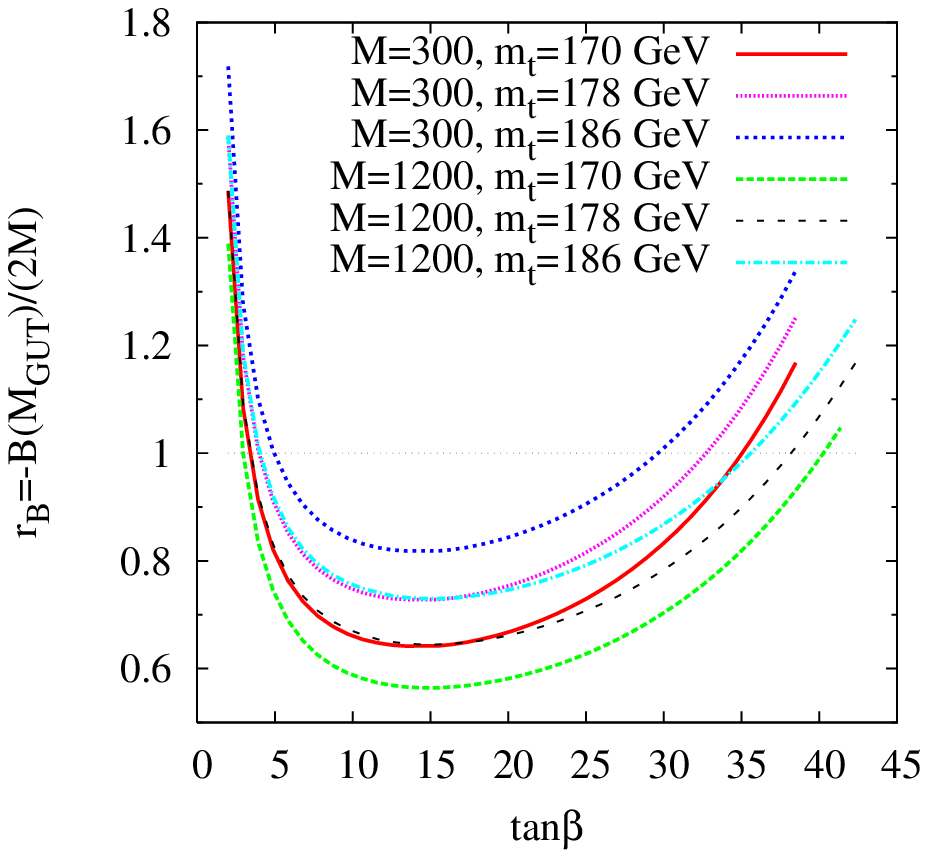}
\caption{Electroweak symmetry breaking constraints 
in the fluxed MSSM for (a) $\mu>0$ and (b)
  $\mu < 0$ and different values of $M$ and $m_t$.
The horizontal line is the naive prediction of the model. The curves end when
  there is no viable model (see text). In (a), points are plotted where 
  models exist which break electroweak symmetry correctly, but which 
possess a charged lightest supersymmetric particle (LSP). 
\label{fig:ewsb}}}
Fig.~\ref{fig:ewsb} shows the prediction for $B(M_{GUT})$
 as a function of $\tan \beta$, for $\mu>0$ and $\mu<0$.
 The prediction is written in terms of $r_B=B(M_{GUT})/(-2M)$.
 As eqs.~(\ref{bCond}), (\ref{xtraCond}) state, 
 the model predicts $r_B=1$, which is indicated by a
 dotted horizontal line. We pick two different values of the mass scale $M$
 and three different values of $m_t$ and show the prediction for $r_B$
 in each case. In Fig.~\ref{fig:ewsb}a we see that the correct value of 
 $r_B$ for $\mu>0$ is impossible to achieve for $m_t=178\pm 4$ GeV within
 2$\sigma$ of its  central value and for $M=300$ GeV or $M=1200$ GeV. We have
 checked numerically that this  result holds for all $M>100$ GeV. 
The curves are truncated on the right hand side of the plot when $m_{{\tilde
 \tau}_1} < M_{\chi_1^0}$, which is ruled out cosmologically for a stable
 neutralino lightest supersymmetric particle (LSP). For completeness,
 we also show extra points where such a condition is violated. 
The plot shows that
 significant departures from the $r_B=1$ prediction are required for the model
 to become viable (for example, $r_B=0.85$ can work for $m_t=186$ GeV). 
Fig.~\ref{fig:ewsb}b, on the other hand, shows that EWSB is viable
 for  $\mu<0$ and that there are two possible solutions for $\tan \beta$, one
 around 3-5 and one at higher $\tan \beta>25$. This result holds for other
 values of $M$.
The curves are truncated on the right hand side when
 $m_A^2<0$, indicating an unsuccessful electroweak minimum. 
In summary, the plots in Fig.~\ref{fig:ewsb} tell us that the model 
 selects a particular sign of $\mu$ and two specific ranges of $\tan \beta$.

Some features in Fig.~\ref{fig:ewsb} can be qualitatively understood
in terms of the minimisation conditions and RGE effects.
For instance, we can write $B(M_{GUT})$ as a sum of two parts, 
$B(M_{GUT}) =  B(M_{SUSY}) + \Delta B$,
where $B(M_{SUSY})$ is determined by eqs.~(\ref{mucond}), (\ref{Bcond}),
and $\Delta B$ encodes the RGE evolution
between $M_{SUSY}$ and $M_{GUT}$, which reads
\begin{equation}
\Delta B \simeq \frac{1}{8 \pi^2}\int_{M_{SUSY}}^{M_{GUT}} 
( 3  h_t^2 A_t +  3  h_b^2 A_b + h_\tau^2 A_\tau 
+ g'^2 M_1 +  3 g^2 M_2) \;  d \ln Q
\end{equation}
at one-loop order.
The $\Delta B$ contribution to $B(M_{GUT})$ is always significant,
and increases at large $\tan\beta$ because then the terms proportional
to $h_b^2, h_\tau^2$ increase.
On the other hand, eq.~(\ref{Bcond}) tells us that $B(M_{SUSY})$ 
is sizable at small $\tan\beta$ only, and that it flips sign if 
we switch from $\mu >0$ to $\mu <0$. This explains why 
Fig.~\ref{fig:ewsb}a and  Fig.~\ref{fig:ewsb}b
are so different at small $\tan\beta$. 
At large $\tan\beta$ the plots are more similar to each other
because $B(M_{SUSY})$ is suppressed, so the main contribution 
to $B(M_{GUT})$ is $\Delta B$, which is less sensitive
to $sgn(\mu)$. The dependence of $\Delta B$ on $sgn(\mu)$
is mainly induced by $h_b, h_\tau$. These couplings are not
exactly the same for $\mu >0$ and $\mu <0$, because their
relation with the fermion masses $m_b, m_\tau$ is affected 
by threshold corrections proportional to $\mu \tan\beta$.
As a result, Fig.~\ref{fig:ewsb}a and Fig.~\ref{fig:ewsb}b
exhibit some differences also at large $\tan\beta$.

Also the fact that the curves in Fig.~\ref{fig:ewsb}a end
before the corresponding curves in  Fig.~\ref{fig:ewsb}b
is mainly a consequence of the different values of
$h_b, h_\tau$ for positive or negative $\mu$. 
As we have mentioned above, the curves end because
either the lighter stau mass or the pseudo-scalar Higgs mass
becomes too small (eventually tachyonic). 
The behaviour of such masses is shown in Fig.~\ref{fig:masses},
for $\mu$ either positive or negative.
Since the lightest stau is mainly ${\tilde \tau}_R$,
the behaviour of $m^2_{{\tilde \tau}_1}$ reflects that of
$m^2_{{\tilde \tau}_R}$, as we have checked. 
The latter mass is $M^2$ at $M_{GUT}$. At smaller scales, 
$m^2_{{\tilde \tau}_R}$ receives negative RGE corrections 
proportional to $h_\tau^2$, which are sizable at large $\tan\beta$ and
eventually drive  $m^2_{{\tilde \tau}_R}$ to negative values.
Stau mixing (proportional to $\tan \beta$) also decreases $m_{{\tilde
    \tau}_1}$,  
but this is sub-leading to the RGE effect for the allowed regions of 
parameter space.
As regards $m_A^2$, it is useful to recall its tree-level expression
at large $\tan\beta$, which is $m_A^2 \sim m_{H_1}^2 - m_{H_2}^2 -M_Z^2$. 
In fact, we have checked that the behaviour of $m_A^2$ in  
Fig.~\ref{fig:masses} reflects that of $m_{H_1}^2 - m_{H_2}^2$.
The latter quantity is driven to small values 
at large $\tan\beta$ because of RGE corrections 
proportional to $h_b^2$ or $h_\tau^2$.

\FIGURE{\twographs{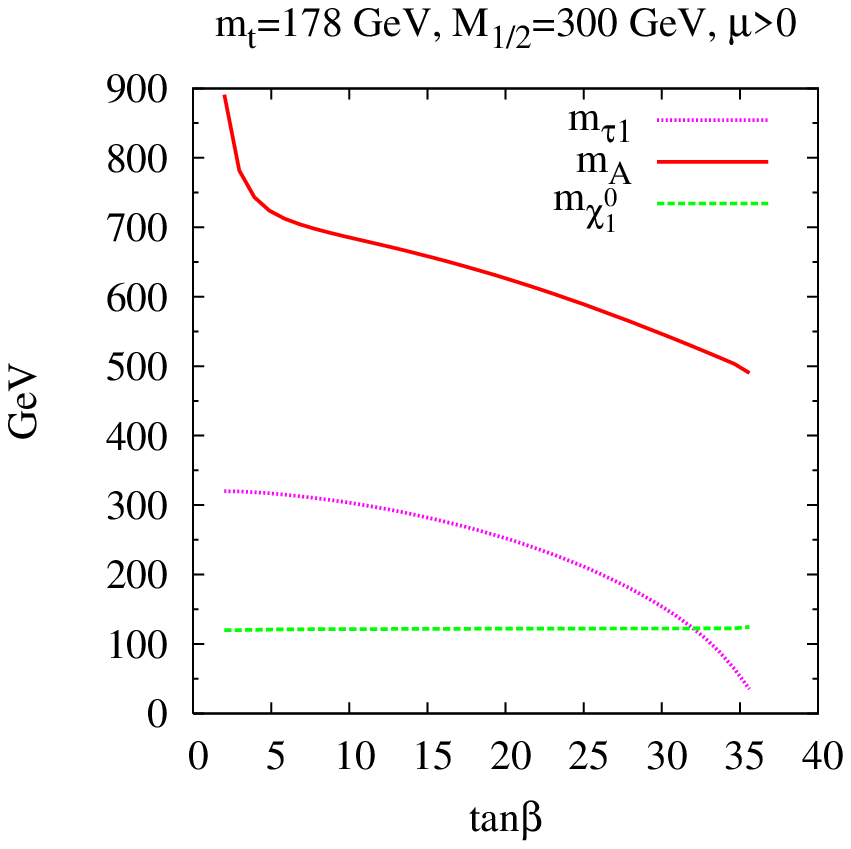}{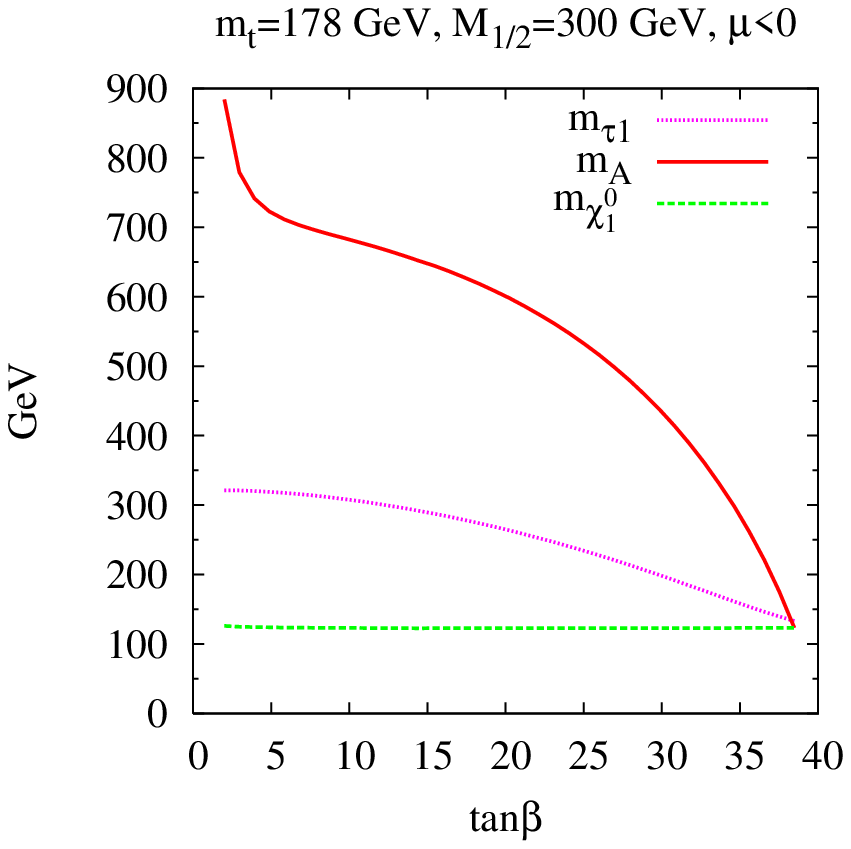}
\caption{Mass variation of the lightest stau, the pseudo-scalar Higgs and the
  lightest neutralino with $\tan \beta$
for $M=300$ GeV, $m_t=178$ GeV and (a) $\mu>0$, (b)
  $\mu<0$.}\label{fig:masses}}

The test of proper EWSB, which we have discussed above, 
is a crucial one for the viability of the model.
Indeed, it is necessary for the existence of a phenomenologically realistic 
local minimum. Although this property is sufficient for the 
phenomenological discussions in the next sections, for completeness 
we should add that deeper minima seem to exist, with large
Higgs vacuum expectation values (VEVs) and broken colour and/or charge. 
In particular, we have examined some field directions
studied in \cite{Komatsu:1988mt}, \cite{Casas:1995pd}, which 
involve the Higgs doublet $H_2$, the selectron doublet $L_e$ and 
a left-right pair of sbottom (or stau) fields.
For sufficiently large field values, $|H_2| \simeq |L_e|
\gg M/h_j $ (with $h_j=h_b$ or $h_\tau$), the leading term 
of the potential along such a direction is 
$V \sim (m_{H_2}^2 + m_{L_e}^2)|H_2|^2$,
where the soft masses are evaluated at a renormalisation
scale $Q \sim |H_2|$. That sum of soft masses is positive 
at $M_{GUT}$, then it decreases and eventually becomes negative,
because of the behaviour of $m_{H_2}^2$. 
This sign flip occurs at quite a large scale 
$Q_*  \sim 10^{9} {\rm GeV} \gg M$, 
hence the potential develops a very deep negative minimum in which
$|H_2| \sim Q_*$ and  $|V| \sim M^2 Q_*^2 \gg M^4 $.
The tunnelling rate between the realistic minimum and such a minimum is
exponentially suppressed by a factor $e^{-S}$, where $S$ is proportional to
$1/h_b^2$ (or $1/h_{\tau}^2$) multiplied by a large numerical coefficient
(see {\it e.g.} \cite{strum}).
We will assume that this suppression is sufficient such that the 
probability of the universe to have tunnelled into the wrong minimum is small.
In any case, a full computation of the 
tunnelling rate goes well beyond the scope of this paper. 
Note that the situation concerning charge/colour-breaking 
minima is similar to that in the case of  another popular
string-inspired scheme, dilaton dominated
SUSY-breaking. Also in this  case lower minima other than the 
standard EWSB minimum appear \cite{clm} and the assumption that we
live in a metastable local vacuum provides a natural way out.

In what follows, we use eq.~(\ref{bCond}) (or eq.~(\ref{xtraCond}))  
as a boundary condition on $B(M_{GUT})$. We then {\em predict} 
$\tan \beta$ using 
two different numerical methods. If one imposes the $\tan
\beta$ prediction coming from eqs.~(\ref{mucond}),~(\ref{Bcond}), 
it turns out that one always
obtains the lower $\tan \beta$ solution of EWSB. We therefore use this
strategy when studying the low $\tan \beta$ solution. For the high $\tan
\beta$ solution however, we revert to the default {\tt \small SOFTSUSY}
calculation, determining the correct value of input $\tan \beta$ by the
method of bisection, using the fact that the curves in Fig.~\ref{fig:ewsb}b 
are monotonic near the high $\tan \beta$ solution. 

\section{Phenomenology}

We now detail the phenomenological constraints that we will place upon the
model. We impose constraints coming from the decay $b \rightarrow s \gamma$ by
calculating its  value with the aid of {\small \tt
  micrOMEGAs1.3.1}~\cite{Belanger:2004yn,Belanger:2001fz} linked to {\small
  \tt micrOMEGAS} via the SUSY Les Houches Accord~\cite{Skands:2003cj}.
The experimental value for the branching ratio of the process $b \rightarrow s
\gamma$ is $(3.52 \pm 0.30) \times 10^{-4}$~\cite{exptBsg}. Including
theoretical errors~\cite{Gambino:2004mv}
($0.30 \times 10^{-4}$) coming from its prediction by adding the two
uncertainties in quadrature, we impose  
\begin{equation}
2.3 \times 10^{-4} <BR(b \rightarrow s \gamma)  < 4.8   \times 10^{-4}
\label{bsgcon}
\end{equation}
at the 3$\sigma$ level upon our prediction of $BR(b \rightarrow s \gamma)$.
{\tt \small micrOMEGAs1.3.1} also calculates the SUSY contribution to the
anomalous magnetic moment of the muon, $\delta a_\mu$. The experimental
measurement of $a_\mu =(g_\mu -2)/2$ gives a very precise result, 
$a_\mu^{exp}=(11 659 208 \pm 6) \times 10^{-10} \cite{Bennett:2004pv}$. 
It is difficult to predict the Standard Model value $a_\mu^{SM}$ 
reliably at this level of accuracy. The estimates vary from being 
0.7-3.2$\sigma$ lower than the experimental number (see {\it e.g.}
\cite{Passera:2004bj}). As a guide, we will take for $a_\mu^{SM}$  
the value $(11659189 \pm 6) \times 10^{-10}$~\cite{deTroconiz:2004tr}, 
which is $2.3\sigma$ lower than $a_\mu^{exp}$. 
This would imply that the new physics contribution to $a_\mu$
is subject to the 3$\sigma$ constraint
\begin{equation}
-6 \times 10^{-10 }< \delta a_\mu < 44 \times 10^{-10}. \label{muanom}
\end{equation}
Many authors (see for example~\cite{Chattopadhyay:2001vx}) 
have used the MSSM to explain the discrepancy between the 
experimental determination and the SM prediction of $a_\mu$.
It usually happens in the MSSM that 
$\delta a_\mu$ has the same sign as $\mu$. 
This is true also in the special version of the MSSM we are 
discussing. In particular, since EWSB prefers negative $\mu$, 
$\delta a_\mu$ is predicted to be negative, so the
discrepancy is not alleviated in our scenario.  
Eqs.~(\ref{bsgcon}) and (\ref{muanom}) will prove to be strong 
constraints upon the fluxed $\mu<0$ MSSM boundary conditions. 

The LEP2 collaborations have put stringent limits upon the lightest Higgs
boson in the MSSM~\cite{Barate:2003sz}. In the decoupling limit of $M_A \gg
M_Z$ (applicable to the results discussed here where we apply limits to the
lightest CP even Higgs boson), the 3$\sigma$ limit upon the mass is
$m_h > 114$ GeV. The error upon the theoretical prediction is estimated to be
typically $\pm 3$ GeV for non-extreme MSSM parameters~\cite{Allanach:2004rh}
and so the bound 
\begin{equation}
m_h > 111 \mbox{~GeV}
\end{equation}
is imposed upon the {\small \tt SOFTSUSY1.9} prediction of $m_h$.

We can also test the hypothesis 
that the lightest neutralino is a stable particle by examining its dark matter
properties. We will compare the 
prediction of relic dark matter density from thermal production in the early
universe by {\tt micrOMEGAs} with the 3 $\sigma$
WMAP~\cite{Spergel:2003cb,Bennett:2003bz} constraint upon the relic density:
\begin{equation}
0.084 <\Omega_{DM} h^2 < 0.138.
\label{omdm}
\end{equation}
It is important to realise that the prediction of {\tt micrOMEGAs} and the 
constraint from WMAP are made in the context of a `standard' cosmological 
model ($\Lambda-$CDM). In the rest of the present paper, we will assume 
that the $\Lambda$-CDM model describes cosmology well. Then, any
prediction lower than the range in eq.~(\ref{omdm}) would require 
additional forms of dark
matter or non-thermal production and any prediction greater than that range
would require the neutralino to be unstable  (either if the neutralino
were not the LSP or if R-parity were violated, for example) and
for some other particle to constitute the dark matter.

\subsection{Low $\tan \beta$ branch of $\mu<0$}

Fig.~\ref{fig:flavConNeg}a shows the constraints coming from $b\rightarrow s
\gamma$ and $\delta a_\mu$ for the low $\tan \beta$ branch of solutions to the
fluxed SUSY breaking boundary conditions as a function of $M$, the free
SUSY breaking mass scale.
\FIGURE{\twographsg{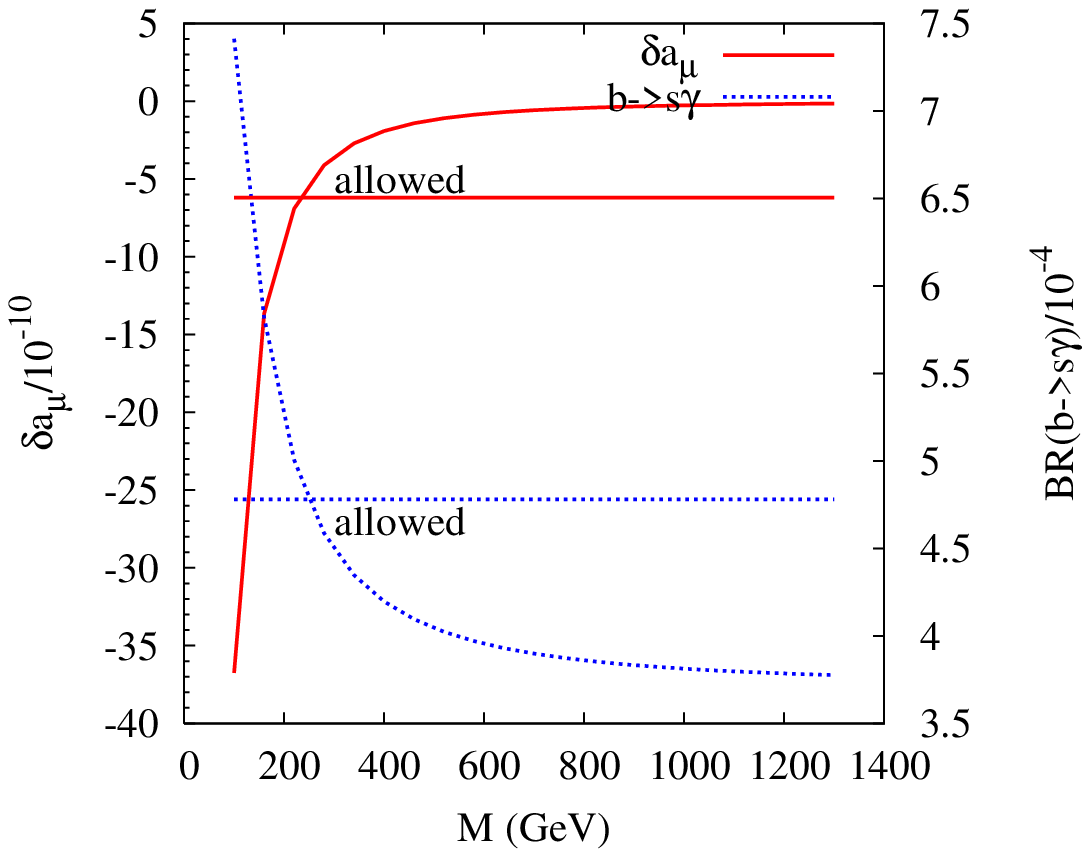}{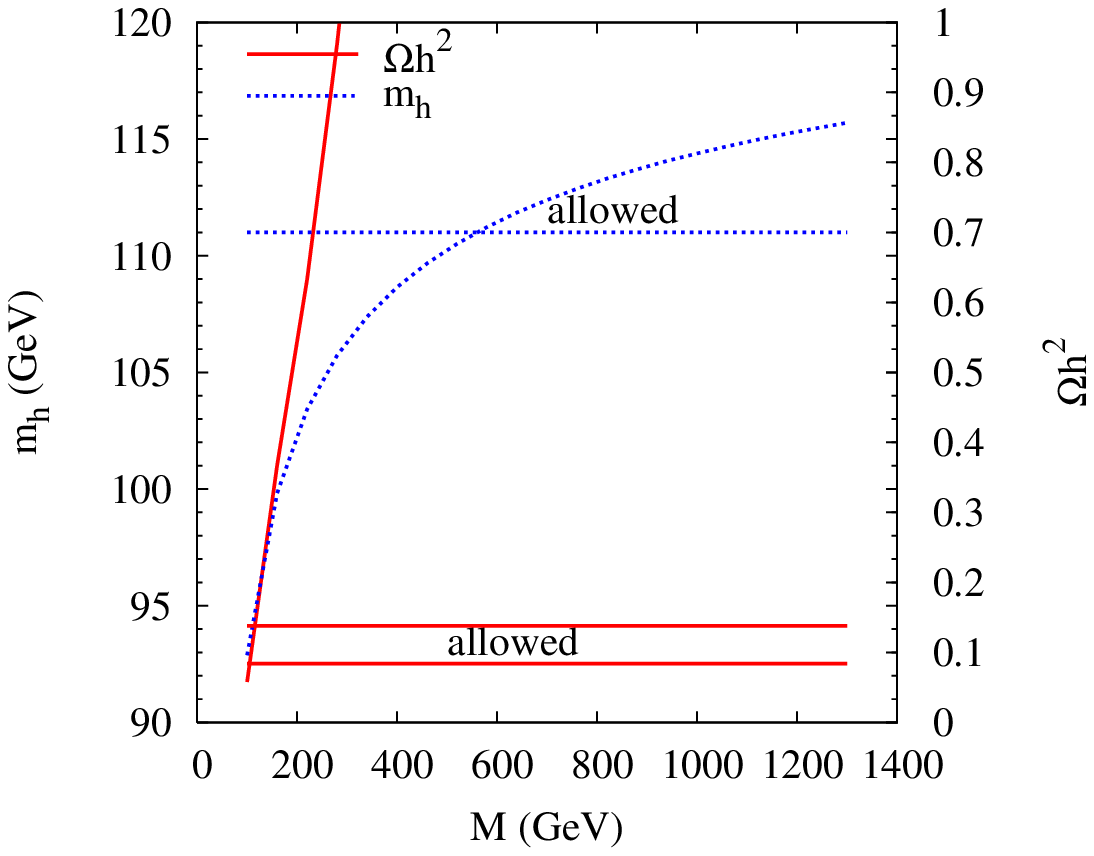}
\caption{Constraints on the low $\tan \beta$ branch
  coming from (a) BR($b \rightarrow s \gamma$) and the
  anomalous magnetic moment of the muon, (b) lightest CP-even Higgs mass and
  dark matter.  The horizontal lines display (a)
  the lower bound on $\delta a_\mu$ and the upper
  bound on BR($b \rightarrow s \gamma$), (b) the lower bound on $m_h$ coming
  from LEP2 constraints and the allowed region coming from the assumption that
  the neutralino makes up the entire dark matter relic density of the
  universe. \label{fig:flavConNeg}}}
We see from Fig.~\ref{fig:flavConNeg}a that the bounds coming from the
anomalous magnetic moment of the muon and $b\rightarrow s \gamma$ have a
  similar effect and imply that 
$M>270$ GeV. From Fig.~\ref{fig:flavConNeg}b, we observe that $M>550$ GeV is
required by the constraints upon the lightest Higgs mass $m_h$. This
last bound is notoriously sensitive to the value of $m_t$ 
(see e.g.~\cite{Allanach:2004rh}) and will change significantly 
if we depart from
  the default central value of $m_t=178$ GeV. We will investigate this effect
  below, but we advance that the bound $M>550$ is substantially
relaxed and values $M>300$ GeV are allowed for $m_t$ values 2 $\sigma $
above the central value $m_t=178$ GeV. 
The prediction of $\Omega h^2$ shows that the neutralino is {\em not}
  compatible with being stable dark matter in the low $\tan \beta$ branch. 
It is only compatible with the WMAP constraint (shown by the region between
  two horizontal lines
  on the figure) for $M \approx 100$ GeV, where the model is already excluded
  by $m_h$, $b \rightarrow s \gamma$ and $\delta a_\mu$.

\FIGURE{\twographs{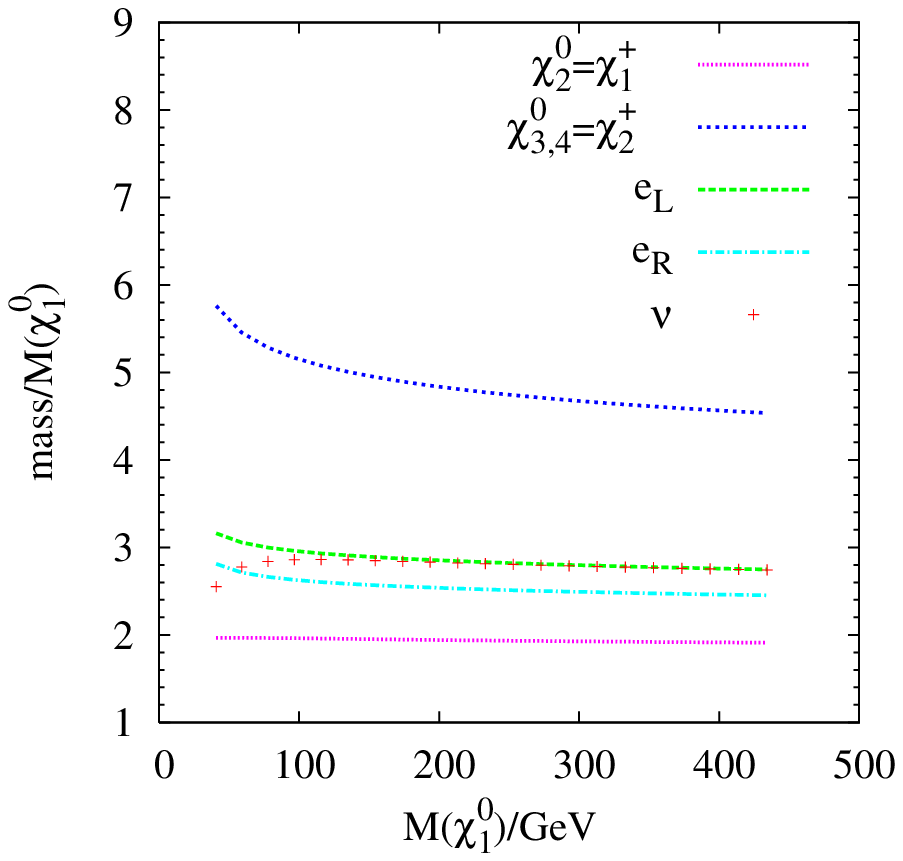}{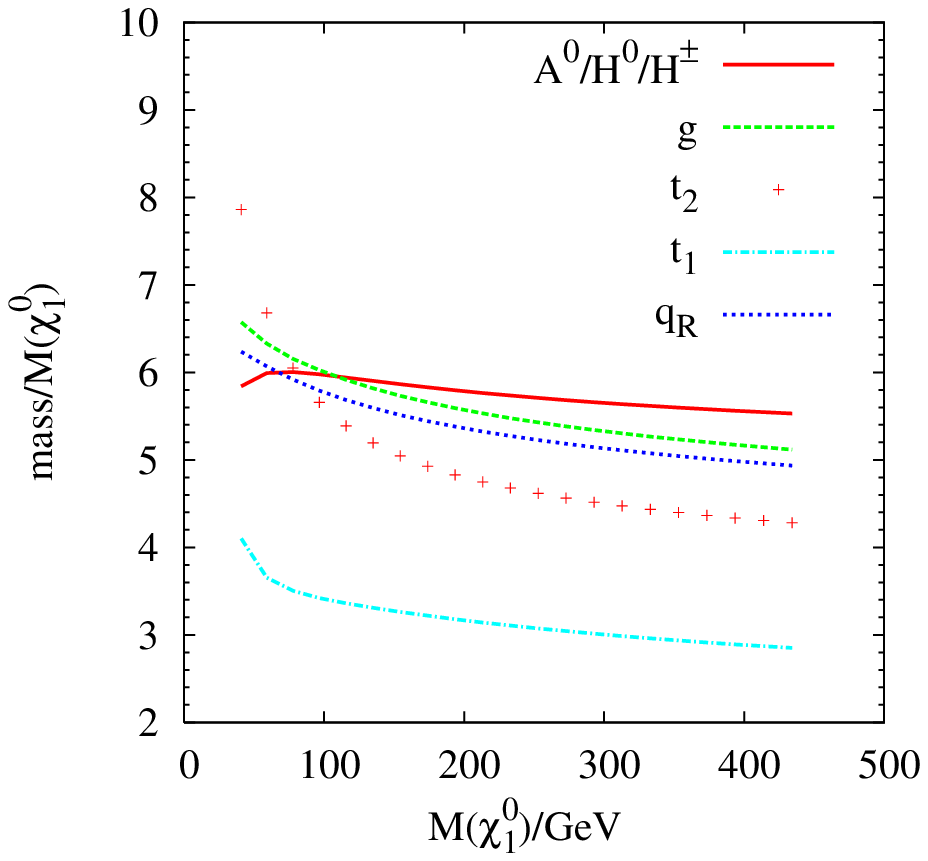}
\caption{MSSM spectrum in the low $\tan
    \beta$ branch. The horizontal range corresponds to $100$ GeV$<M<1000$ GeV.
\label{fig:specLowNeg}}}
The sparticle spectrum of the low $\tan \beta$ branch is shown in terms of
$M_{\chi_1^0}$ in Fig.~\ref{fig:specLowNeg}. The lightest neutralino
$\chi_1^0$ is mainly bino, hence $M_{\chi_1^0} \sim 0.4 M$.
From the figure, we see that 
no other sparticles are particularly close in mass to the LSP, the nearest
being the second-lightest neutralino and the lightest chargino. These two
particles are quasi-degenerate (they are mainly winos)
and are plotted as one curve on the
plot, since individual curves could not be distinguished by eye. 
Also the second lightest chargino and the third and fourth
lightest neutralinos are quasi-degenerate (they are
mainly Higgsinos). In the slepton sector, the masses shown
correspond to $\tilde{e}_L$, $\tilde{e}_R$, $\tilde{\nu}_e$,
but the same (or almost the same) results hold for the 
other generations. As regards 1st/2nd generation squarks, 
the left-handed ones (not shown) are almost degenerate with the gluino,
whereas the right-handed ones ($\tilde{q}_R$) are slightly lighter.
In the third generation sector we only show the stop squarks.
The heavier sbottom, which is mainly $\tilde{b}_R$, is close 
to $\tilde{q}_R$. The lighter sbottom, which is mainly $\tilde{b}_L$,
is close to the heavier stop ($\tilde{t}_2$), except for small 
$M_{\chi_1^0}$. The mass ratios do not vary much with increasing $M$ 
except for the heavier stop. If the absolute masses were plotted, 
they would increase linearly with $M$ (approximately).  

\subsection{High $\tan \beta$ branch of $\mu<0$}

We now turn to the other branch of higher $\tan \beta$ that is valid for
$\mu<0$. 
\FIGURE{\twographsg{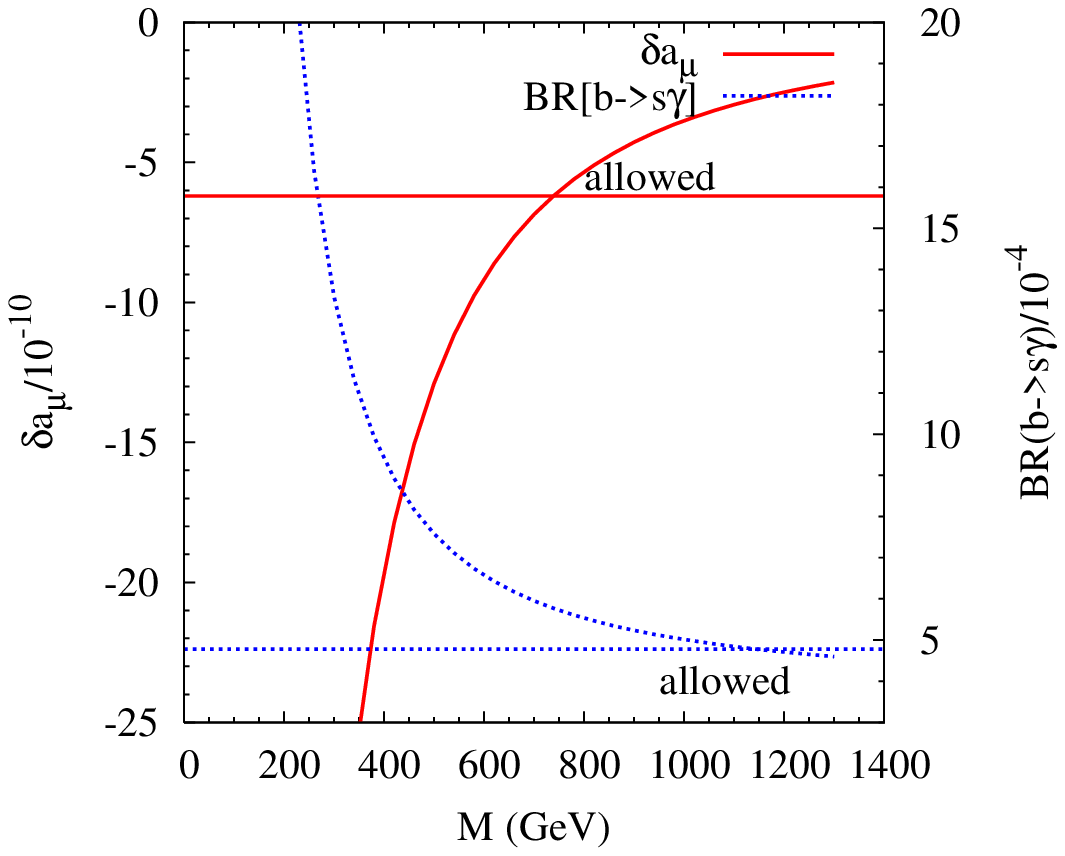}{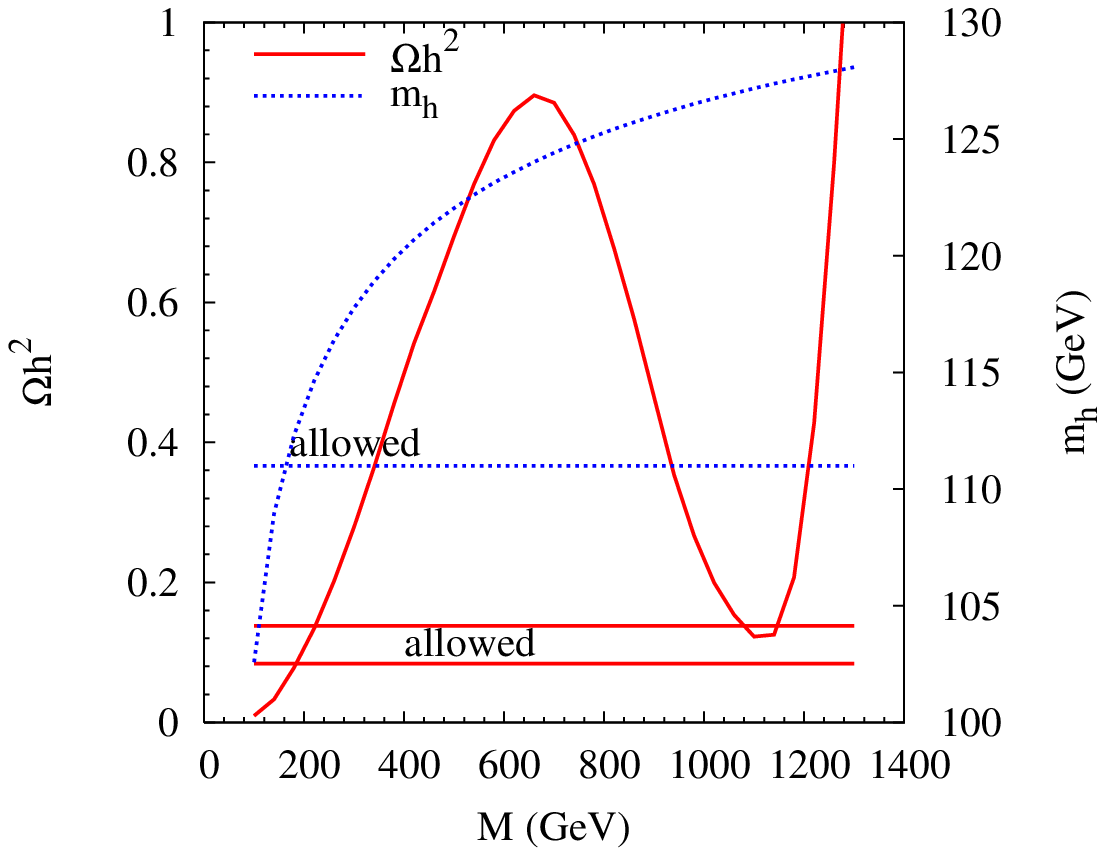}
\caption{Constraints on the high $\tan \beta$, $\mu<0$ branch. 
(a) constraints coming from the anomalous magnetic moment (with the horizontal
  line showing a lower bound) and from $BR(b \rightarrow s \gamma)$ (with the
  associated horizontal line showing an upper bound). 
(b) constraints from the
  dark matter relic density and the lightest CP-even Higgs mass.
  Horizontal lines show the lower bound on $m_h$ from the LEP2 direct searches
  and the WMAP constraints upon $\Omega h^2$. 
\label{fig:hiNegFlav}}}
Fig.~\ref{fig:hiNegFlav}a shows that, contrary to the low $\tan \beta$
solution, $\delta a_\mu$ and $b \rightarrow s \gamma$ provide much more
stringent constraints: $M>750$ and $M>1100$ GeV respectively.
This is expected, of course, because the SUSY contributions to 
both $\delta a_\mu$ and the $b \rightarrow s \gamma$ amplitude
grow with $\tan \beta$, hence a larger $M$ is needed to provide
the necessary suppression.
On the other hand, Fig.~\ref{fig:hiNegFlav}b shows that $m_h$ provides a much
less stringent constraint of $M>140$ GeV compared to the low $\tan \beta$
case. The main reason is that the tree-level contribution to $m_h$,
which is $\sim M_Z |\cos 2\beta|$, increases for increasing $\tan \beta$. 
The neutralino is compatible with being a stable dark matter candidate
for $M \approx 200$ GeV (which is already ruled out by the $b \rightarrow s
\gamma$ and $\delta a_\mu$ constraints) or $M \approx 1100$ GeV, 
a remarkable point with a heavy MSSM spectrum that passes all constraints. 

\FIGURE{\twographs{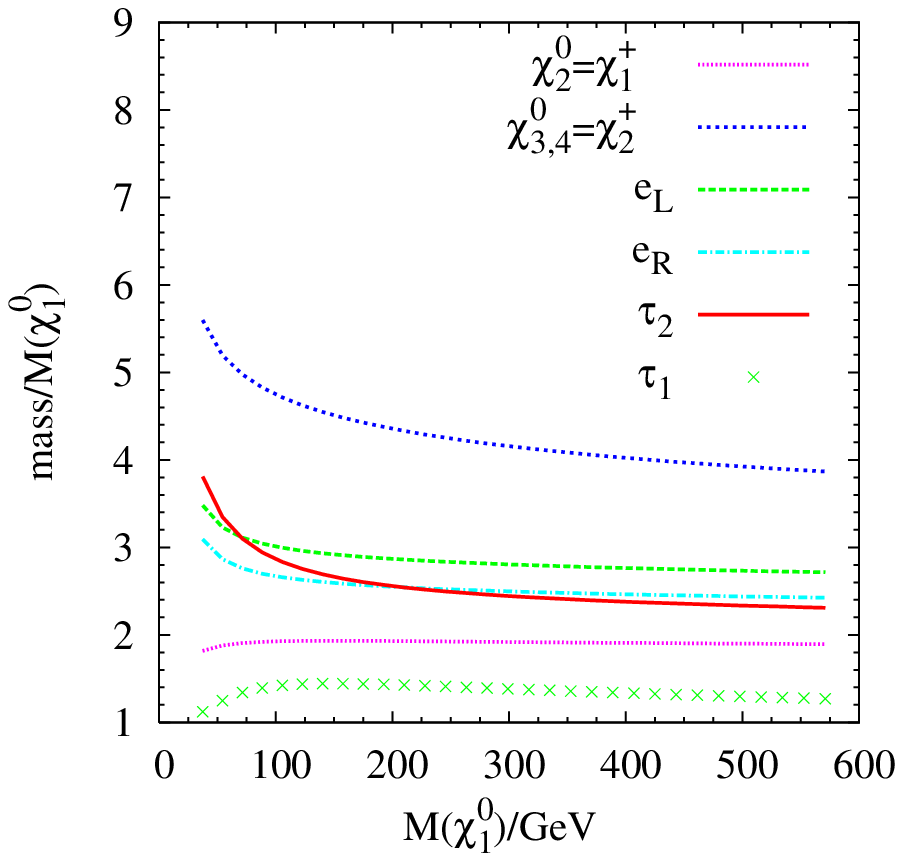}{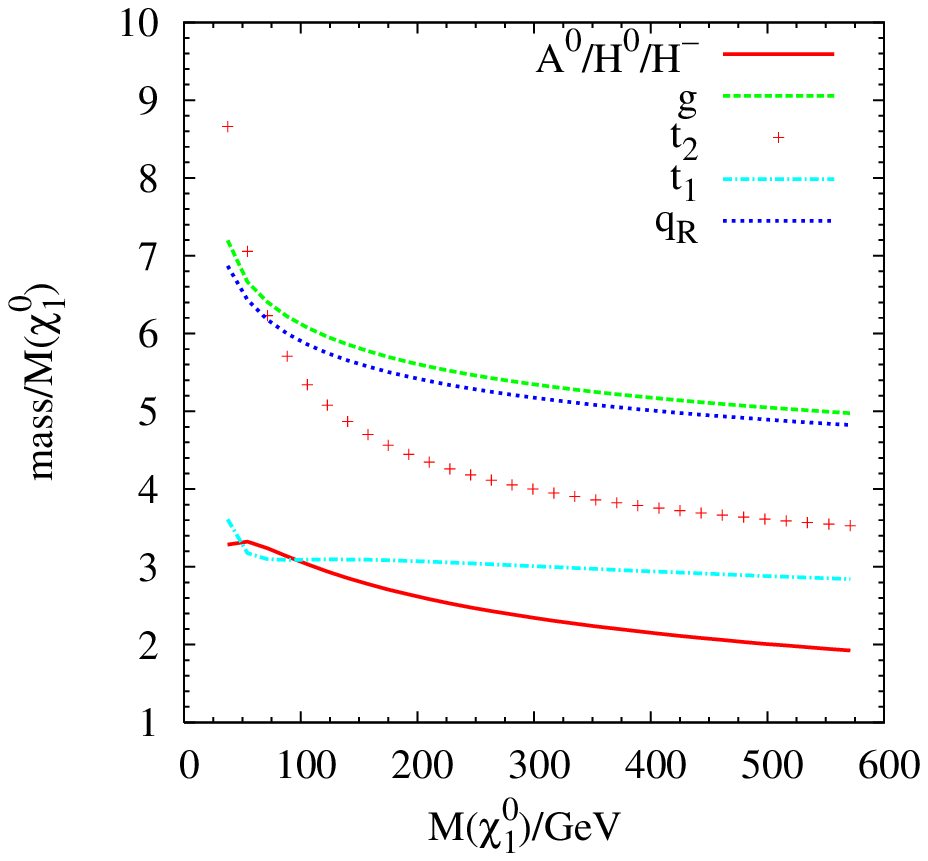}
\caption{MSSM spectrum in the high $\tan \beta$ branch. The horizontal range
  corresponds to $100$ GeV$<M<1300$ GeV. 
\label{fig:specHiNeg}}}
Fig.~\ref{fig:specHiNeg} displays the sparticle spectrum in terms of
$M_{\chi_1^0}$ (which is about $0.4 M$) for the high $\tan \beta$ case. 
Some of the masses (neutralinos, charginos, gluino, 1st/2nd generation
sfermions) are not very different from those in the low $\tan \beta$ 
branch. Some other masses exhibit significant differences.
In particular, each stau is lighter than the corresponding 
selectron\footnote{Sneutrino masses, which are not shown, are 
close to those of the corresponding 
charged sleptons ($\tilde{\nu}_e \sim \tilde{e}_L$ and 
$\tilde{\nu}_{\tau} \sim \tilde{\tau}_2$), except at small 
$M_{\chi_1^0}$.}. Both sbottom squarks (not shown) have masses
between $\tilde{t}_1$ and $\tilde{t}_2$. 
Finally, the heavy Higgs bosons are much lighter than gluinos 
and squarks, in contrast to the previous case.
In particular, at higher values of $M_{\chi_1^0}$ the value of $m_A$ 
approaches $2 M_{\chi_1^0}$. This  enhances the 
$\chi_1^0 \chi_1^0 \rightarrow A^0 \rightarrow b \bar{b}$ or $\tau
\bar{\tau}$ annihilation channel through the pseudo-scalar Higgs pole,
explaining the dip in $\Omega h^2$ displayed in Fig.5-b.

As emphasised above, for $M \approx 1100$ GeV the lightest 
neutralino provides the appropriate dark matter density and
all other experimental constraints are fulfilled at the 
same time. The corresponding spectrum is quite heavy,
with $\chi_1^0$ (the LSP) at around 500 GeV,
${\tilde \tau}_1$ around 600 GeV, $\chi_2^0$ and $\chi_1^+$
around 900 GeV, the heavy Higgs bosons around 1 TeV,
and all other sparticles heavier than 1 TeV.
This scenario could pose some problems on the experimental side,
but SUSY should still be detectable at the LHC with sufficient
luminosity.
On the theory side, a possible draw-back is that such high $M$ seems 
to require significant fine-tuning in EWSB, since $M_Z^2 \ll M^2$.
For instance, proper EWSB requires $|m_2^2| \ll M^2$ at low energy, where 
$m_2^2= m_{H_2}^2 + \mu^2$ is the coefficient of $|H_2|^2$.
Hence an accurate cancellation is needed between the low-energy values 
of $m_{H_2}^2$ and $\mu^2$, which are a priori unrelated parameters
with values ${\cal O}(M^2)$.
On the other hand it may be that some fundamental model predicts 
a relationship between $\mu$ and the soft SUSY breaking parameters at the
unification scale which leads to the appropriate cancellation at low energy.
In such a scenario the fine-tuning problem would be alleviated,
or even removed. In this respect, it is interesting to note   
that the values of $\mu(M_{GUT})$ required
by proper EWSB are numerically close to $-2 M$, in our case. 
This is shown in Fig.~\ref{fig:mu} for both $\tan \beta$ branches.
\FIGURE{\epsfig{file=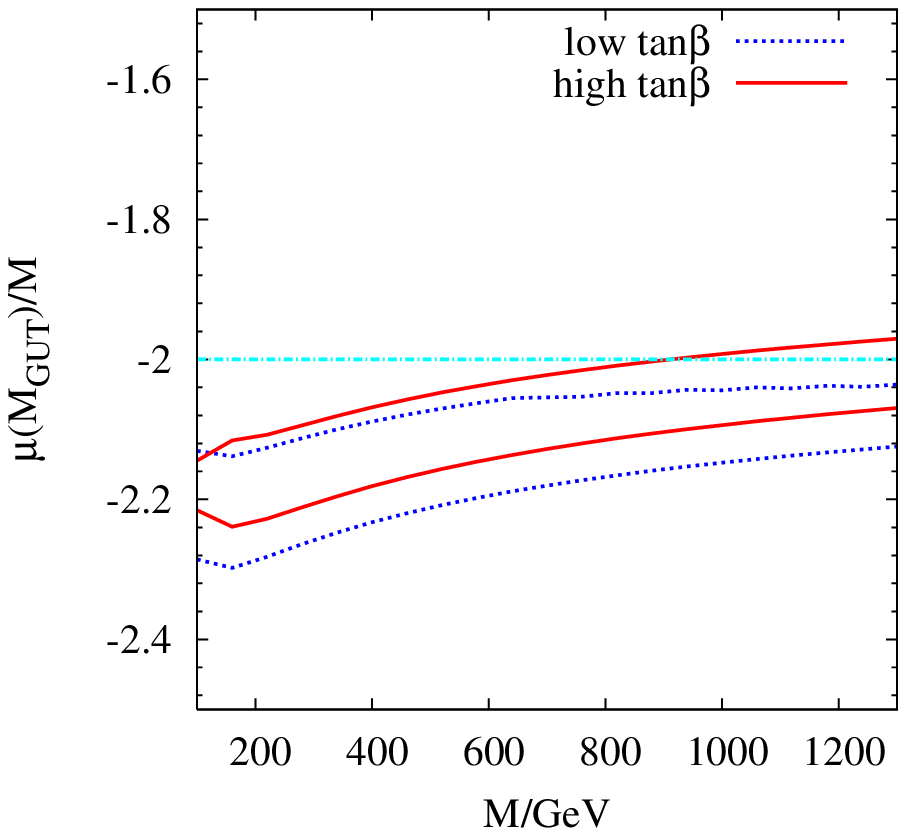, width=\wth}
\caption{The ratio $\mu(M_{GUT})/M$ as required by EWSB, for
the high and low $\tan \beta$ branches. In each case,
the band is obtained by varying $m_t$ in its $2 \sigma$ 
range. The horizontal line corresponds to $\mu(M_{GUT})/M=-2$.
\label{fig:mu}}}
In particular, we can see that a hypothetical relation
$\mu(M_{GUT})=-2M$ is consistent with EWSB in the high $\tan \beta$ 
branch. It is also interesting that this happens for large $M$ only
($M > 900$ GeV), which is compatible with the experimental 
constraints discussed above.
So, if such a relationship was found to hold due to some underlying 
physics at the unification  scale\footnote{In fact it turns out 
that in simple toroidal orientifold
Type IIB compactifications $\mu $ and $M$ correspond to different
types of 3-form fluxes and in some simple cases they are related due
to constraints coming from certain  tadpole cancellations
(see e.g.\cite{font,msw}). So it is not inconceivable that further 
constraints like $|\mu| = 2 M$ could appear in some specific 
situation.}, one would have a model in which a little hierarchy
$M_Z/M \sim 0.1$ could emerge without requiring a special fine-tuning
among mass parameters\protect\footnote{Only dimensionless parameters, 
i.e. gauge and Yukawa couplings, would need to be adjusted. 
This tuning, however, may be regarded as a separate problem.}. 
In such a model, in other words, the simple relationships 
among MSSM mass parameters at the unification scale, combined with 
the RG evolution of such parameters, would lead to cancellations 
in the low-energy Higgs potential, so that the smallness of
$M_Z/M$ could be (at least in part) justified.
 
\subsection{$\mu>0$ Case}

We turn now to the $\mu>0$ set of solutions. Since we have found
no solutions for the strict model prediction of $r_B=1$,
we are forced to consider perturbations of the fluxed boundary 
conditions. Fig.~\ref{fig:ewsb}a
demonstrates that higher $m_t$ leads to values of $r_B$ closer to one, 
and that values of $r_B=0.85$ are attainable if $m_t=186$ GeV. 
We have taken these two values in order to study the scenario, but for
purposes of brevity neglect to include the plots in this paper, choosing to
summarise the results in text for $M=100-900$ GeV.
$\tan \beta$ is in the range 26-36, monotonically increasing with $M$.
The LEP2 bound on $m_h$ implies a mild constraint only ($M>150$ GeV).
The  $b \rightarrow s \gamma$ bound implies a stronger  
constraint ($M>500$ GeV), due to the high value for $\tan \beta$.
This bound, however, is weaker than the bound that holds for $\mu<0$, 
because for $\mu>0$ some cancellations take place in the 
$b \rightarrow s \gamma$ amplitude.
The advantage of $\mu>0$ is even more evident in the case of 
$\delta a_\mu$, which is positive, so that the predicted $a_\mu$ 
more easily agrees with the measured value. In fact, the constraint 
from $\delta a_\mu$ is not very restrictive ($M>230$ GeV), 
despite the high values for $\tan \beta$.

The model's spectrum looks remarkably similar to Fig.~\ref{fig:specHiNeg}.
The main differences are that the heavy Higgs bosons have larger masses, 
and the lightest stau is closer in mass to the neutralino. 
The latter feature results in two small regions in $M$ where dark matter is
compatible with the WMAP constraint. For the specified inputs, such regions
have $M=160-190$ GeV and $M=640-650$ GeV.
Between these two regions, $\Omega h^2$ is higher than the WMAP constraint,
otherwise it is lower.

\section{Stability of Results}

We will now examine how stable the phenomenology is
with respect to variations in experimental inputs, 
perturbations of the fluxed boundary conditions and
theoretical uncertainties in the mass spectrum prediction.
We will consider both $\tan \beta$ branches of the $\mu<0$ case.
In particular, an important question about the high
$\tan \beta$ branch is whether it is possible to obtain a solution 
to all the constraints with a lighter spectrum than $M \approx 1100$ GeV,
taking into account various relevant uncertainties.

In order to investigate the stability of our predictions, we will vary 
one input or boundary condition at a time, keeping all others at their
defaults. These variations will be:
\begin{itemize}
\item
2 $\sigma$ variations in the input value of $m_t$, i.e. 170 GeV $< m_t <$ 186
GeV. 
\item
2 $\sigma$ variations in the input value of $\alpha_s(M_Z)$, 
i.e. $0.115 <\alpha_s(M_Z)< 0.123$. 
\item
Variations of $M_{SUSY}$ by a factor of two in each direction 
(setting the minimum value to be not smaller than $M_Z$), which should give
an estimate of higher-order theoretical uncertainties
(see {\it e.g}. ref.~\cite{Allanach:2004rh}).
\item
10\% variations in the boundary condition for $M_{1/2}$ 
($0.9 <M_{1/2}/M <1.1$), without changing the
boundary conditions for $m_0, A_0, B$.
\item 
10\% variations in the boundary condition for $B$
(that is, $0.9 < r_B < 1.1$ in  eq.~(\ref{xtraCond})),
without changing the boundary conditions for $M_{1/2},m_0, A_0$.
\end{itemize}
In practice, for a given value of $M$, we scan over 20 values of the parameter
being varied, then plot the maximum and minimum values for the constraints
obtained with that scan. 
Fig.~\ref{fig:var} shows the effect of the above variations
on some of the predictions in the high $\tan \beta$ branch 
(a,b,c) and in the low $\tan \beta$ branch (d).
In each panel, we only show variations that have
non-negligible effects.

\FIGURE{
\fourgraphs{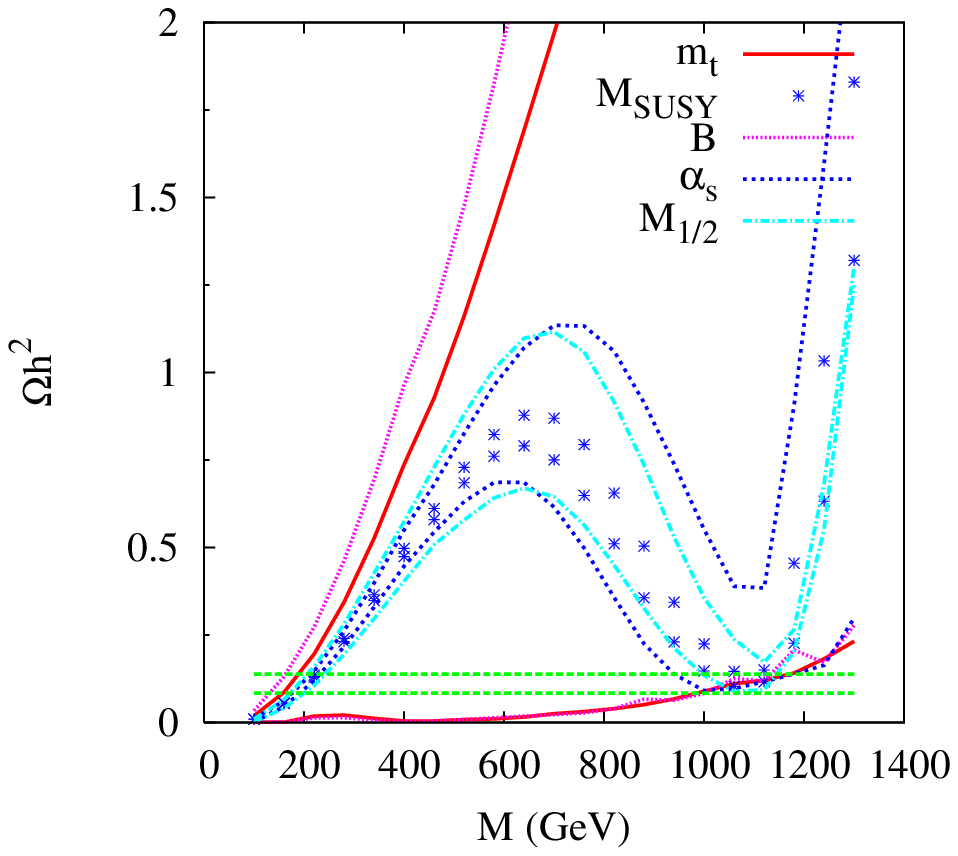}{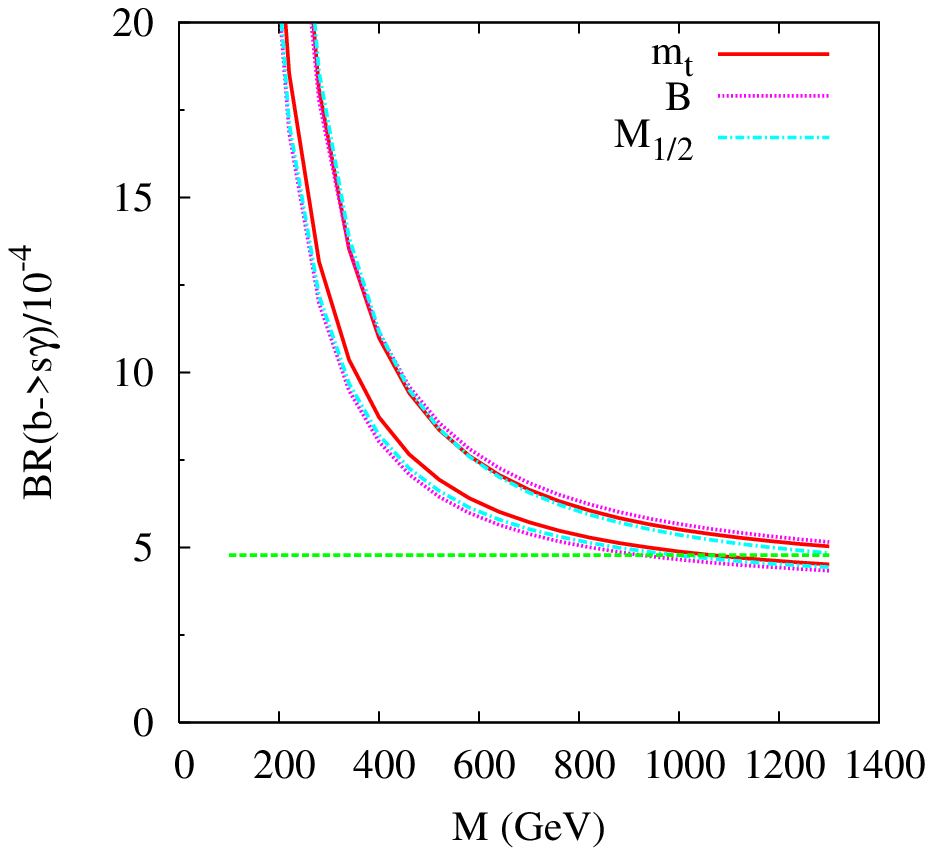}{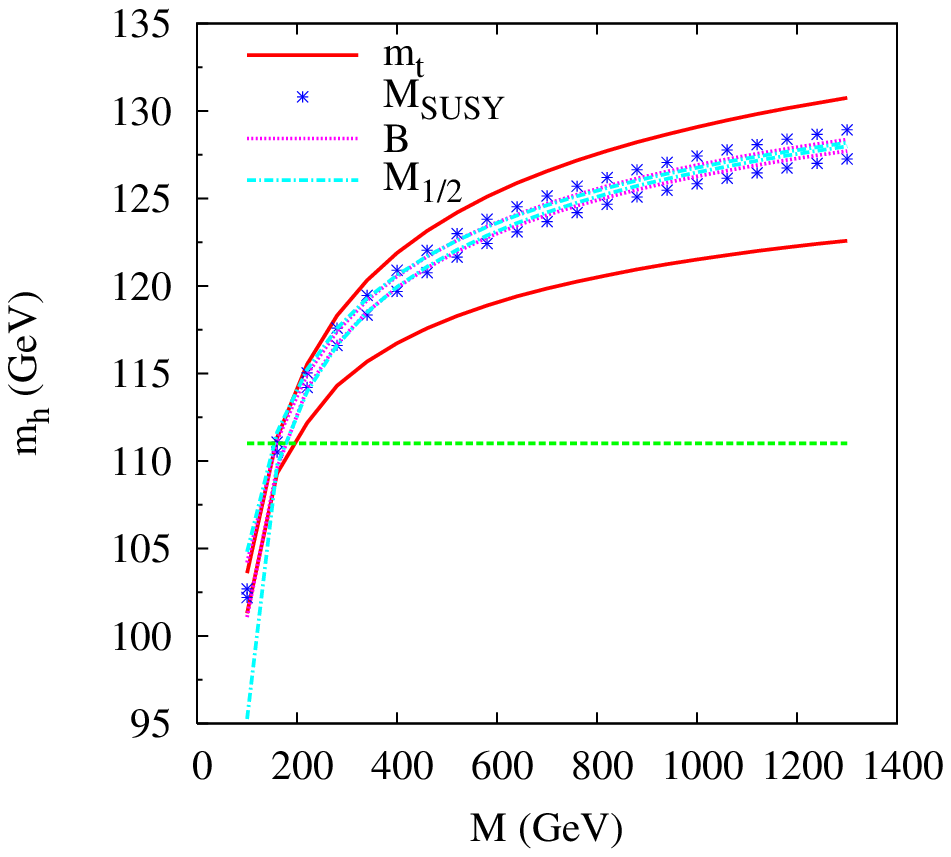}{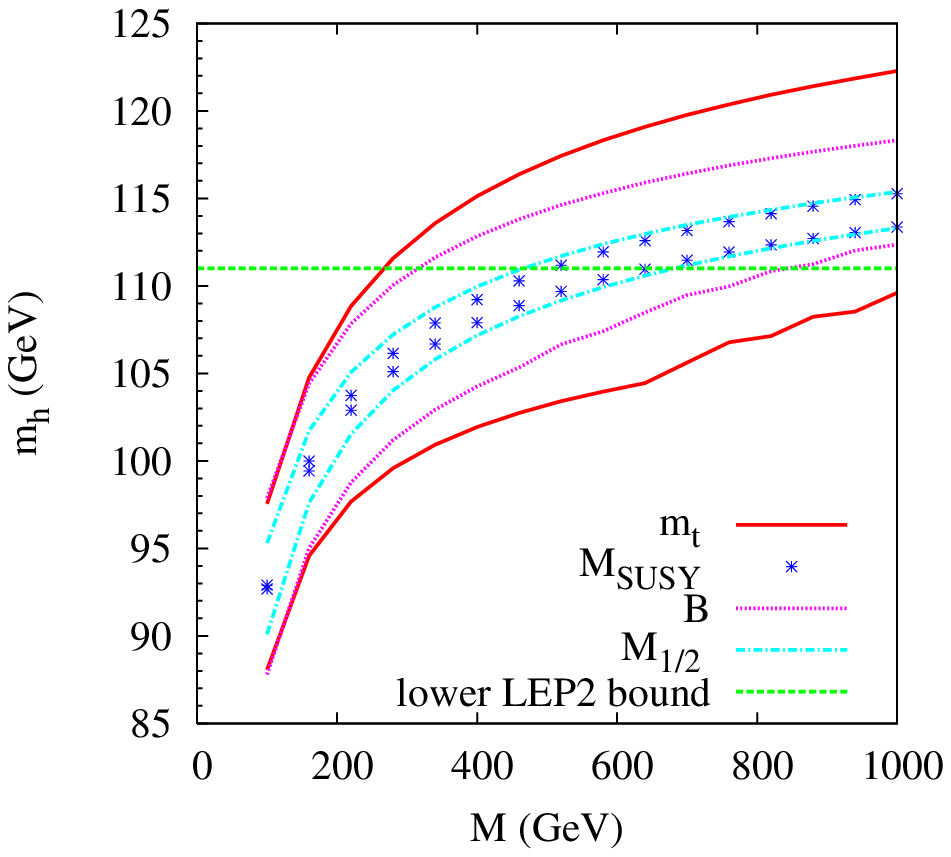}
\caption{Uncertainties in predictions along the $\mu<0$ high 
$\tan \beta$ branch in (a) the dark matter
relic density $\Omega h^2$, (b) $BR(b \rightarrow s \gamma)$
(c) $m_h$. Panel (d) shows $m_h$ in the $\mu<0$ low 
$\tan \beta$ branch. In (a), the region between the two horizontal
lines is favoured by the 3$\sigma$ WMAP constraint for a stable 
neutralino LSP making up all of the dark matter in the universe.
\label{fig:var}}
}

We first consider the high $\tan \beta$ branch. 
Fig.~\ref{fig:var}a shows uncertainties in the relic density prediction. 
The uncertainty under variations of $M_{SUSY}$ is small.
Changing either the $M_{1/2}$ boundary condition by $\pm 10\%$ or
the input value of $\alpha_s(M_Z)$ makes a larger difference, but does
not qualitatively change the region compatible with WMAP, shown by the
region between the 
horizontal lines. However, we see that varying either $m_t$ or the high-scale
boundary condition on $B$ makes any value $200 < M < 1200$ GeV
compatible with the WMAP constraint. 
As explained previously, above $M>200$ GeV, low values of $\Omega h^2$ are
caused when two $\chi_1^0$s annihilate via almost-resonant $s$ channel
pseudo-scalar $A$ Higgs bosons, 
 i.e.\ for $m_A \simeq 2 M_{\chi_1^0}$. 
Since $m_A$ is very sensitive to variations
in $m_t$ or in the $B$ boundary condition, the value of $M$ at which 
the condition $m_A \simeq 2  M_{\chi_1^0}$ is satisfied 
can shift very much under such variations.
Hence the constraint on $M$ from the relic density is 
not so stringent. However, Fig.~\ref{fig:var}b shows that
$b \rightarrow s \gamma$ requires $M$ to be anyhow larger
than about 1 TeV, even if we allow for variations in $m_t$ 
or in the $B$ boundary condition. Thus, it appears that we are stuck with a
heavy spectrum (and the associated large fine-tuning and difficulties for 
detection in colliders) for the $\mu<0$ high $\tan \beta$ branch.
For completeness, we also show the behaviour  
of $m_h$ in Fig.~\ref{fig:var}c, including variations. 
The constraint on $M$ remains quite weak, but the absolute prediction
on $m_h$ has a strong dependence on $m_t$, as usual.

We turn now to the discussion of the low $\tan \beta$ branch.
For default parameters, this branch is not compatible with the stable
neutralino LSP assumption since the relic density predicted is much higher
than the WMAP constraint, as shown previously in Fig.~\ref{fig:flavConNeg}.
Contrary to the case in the high $\tan \beta$ branch, this situation does not
change with any variations, because none of them brings the model 
appreciably closer to a (co-)annihilation region.
If we assume that this problem can be solved by some means (for instance
by assuming that the lightest neutralino is not stable on
cosmological time scales), the strongest bound on $M$ in the 
default calculation
($M>550$ GeV) comes from the LEP2 Higgs mass constraint, 
see Fig.~\ref{fig:flavConNeg}b. 
Fig.~\ref{fig:var}d shows the behaviour of $m_h$ under variations. 
We notice that $m_h$ has the usual strong
dependence on $m_t$, as well as a significant dependence on the
$B$ boundary condition. Indeed, changing the latter parameter
produces a significant relative change in $\tan\beta$,
which in turn affects the tree-level Higgs mass (in this
low $\tan\beta$ branch). 
Fig.~\ref{fig:var}d shows that varying either $m_t$ or the 
$B$ boundary condition results in a lower bound $M>300$ GeV. 
The variations induced by changes in $M_{SUSY}$ or
$M_{1/2}$ have smaller effects upon the bound. 
The bounds coming from the anomalous magnetic moment of the 
muon and from $b\rightarrow s \gamma$ do show smaller variations: 
the bounds on $M$ can vary up to $\sim\pm 10\%$. 

\section{Conclusions}

We have considered the phenomenology of a set of SUSY-breaking 
MSSM  boundary conditions motivated by flux-induced SUSY 
breaking in Type IIB orientifold models. These boundary 
conditions may be interpreted as the Type IIB version of 
modulus-dominance SUSY-breaking considered in the past 
for  heterotic models.
The model is extremely constrained and, after imposing radiative EWSB,
depends on a single parameter, the overall SUSY-breaking scale $M$. 
This is to be contrasted with other string inspired schemes like
heterotic dilaton domination in which, even after imposing EWSB 
conditions, there are still two free parameters.
We find it remarkable that consistent EWSB may be attained 
at all for our  very restricted set of soft terms.

  We have shown that electroweak symmetry breaking is
only compatible with one sign of $\mu<0$, for which there are 
two solutions: one with low  $\tan \beta \sim 3-5$ and one with
high $\tan \beta \sim 25-40$:

\begin{itemize}

\item
The low $\tan \beta$ solution is consistent with having
a SUSY breaking parameter $M>300$ GeV depending on the precise 
value of $m_t$. This may allow for  a relatively light SUSY spectrum. 
On the other hand 
this  low $\tan \beta$ solution is not compatible with the lightest
neutralino constituting the dark matter of the universe, since the predicted
relic density is higher than that implied by WMAP and standard cosmology. 
Thus one would need to have an unstable neutralino, which could be due 
to the existence of an extra lighter SUSY particle or
to the existence of R-parity violating couplings.
In the latter case the collider phenomenology would very much depend 
on the neutralino lifetime.
If it is  much longer than
$\mathcal{O}(10^{-6})$s the lightest 
neutralino will leave detectors intact; collider signatures  
will therefore mimic the usual missing energy R-parity conserving
MSSM signatures and LHC discovery and
measurements~\cite{atlasTDR} should be
easily possible. 
 If the lifetime of the neutralino is shorter than $10^{-6}$s
then the associated collider phenomenology will depend largely on its decay
products.

\item 
The high $\tan \beta$ branch of solutions can
pass all constraints {\em as well as} provide the dark matter density
compatible with WMAP constraints. The spectrum compatible with these
constraints is heavy: with the LSP ($\chi_1^0$) at around 500 GeV 
and all sparticles other than $\chi_2^0, \chi_1^+$ and 
${\tilde \tau}_1$ heavier than 1 TeV. This would 
make sparticle measurements at the LHC practically
impossible. 
With 100 fb$^{-1}$ integrated luminosity, SUSY should still be detectable at
the LHC through the inclusive $E_T^{\mbox{miss}}$
measurement~\cite{Branson:2001pj}, but
more luminosity would be required for other channels involving leptons.

Due to the  heavy spectrum in this high $\tan \beta$ case a certain
amount of fine-tuning of the EWSB conditions would seem to be required.
On the other hand we have argued that if the underlying theory predicted an
approximate relationship $\mu = -2M$ at the unification scale,
such fine-tuning would be substantially alleviated or even removed.

\end{itemize}

\acknowledgments{This work has been partially supported by PPARC.
The work of L.E.I. has been supported in part by the 
Ministerio de Ciencia y Tecnolog\'{\i}a (Spain).
The work of A.B. and L.E.I. has been supported in part
by the European Commission under RTN contracts HPRN-CT-2000-00148
and MRTN-CT-2004-503369.}

\end{document}